\documentclass[aps,prb,floats,showpacs,twocolumn]{revtex4}
\usepackage{graphicx,color}
\pdfoutput=1

\begin{document}
\title{Electron-vibron coupling in suspended carbon nanotube quantum dots}
\author{Eros Mariani}
\author{Felix von Oppen}
\affiliation{Institut f\"ur Theoretische Physik, Freie Universit\"at
Berlin, Arnimallee 14, 14195 Berlin, Germany}
\date{\today}
\begin{abstract}
Motivated by recent experiments, we investigate the electron-vibron coupling in suspended carbon nanotube quantum dots, starting with the electron-phonon coupling of the underlying graphene layer. We show that the coupling strength depends sensitively on the type of vibron and is strongly sample dependent. The coupling strength becomes particularly strong when inhomogeneity-induced electronic quantum dots are  located near regions where the vibronic mode is associated with large strain. Specifically, we find that the longitudinal stretching mode and the radial breathing mode are coupled via the strong deformation potential, while twist modes couple more weakly via a mechanism involving modulation of the electronic hopping amplitudes between carbon sites. A special case are bending modes: for symmetry reasons, their coupling is only quadratic in the vibron coordinate. Our results can explain recent experiments on suspended carbon nanotube quantum dots which exibit vibrational sidebands accompanied by the Franck-Condon blockade with strong electron-vibron coupling. 

\end{abstract}
 \maketitle
\section{Introduction}
\label{sec:introduction}

The coupling of electronic and mechanical degrees of freedom is a promising research avenue in the physics of nanoscale systems.\cite{Roukes,Cleland,Park} 
The detection of electromechanical oscillations in suspended graphene \cite{Bachtold} and carbon nanotubes \cite{Poncharal} represents the current frontier in the field of nanoscale resonators.
Suspended carbon nanotubes (CNT) constitute a particularly interesting model system due to their remarkable electronic properties inherited from the carbon honeycomb structure,  their high mobility as well as their many-faceted mechanical properties such as low dissipation.\cite{Dresselhaus} This is highlighted by several recent experiments on suspended CNT quantum dots which observe pronounced vibrational effects in electronic transport.\cite{LeRoy,Sapmaz,Ilani,Huttel1,Huttel2,Leturcq} These experiments focus on vibrational sidebands in the Coulomb blockade regime and suggest the existence of rather strong but sample dependent electron-vibron coupling, at least for certain vibron modes. Indeed, of the many vibrational modes of CNT, only the radial breathing mode \cite{LeRoy} and the longitudinal stretching mode \cite{Sapmaz,Huttel1,Huttel2,Leturcq} have been observed in transport experiments.
The electron-vibron coupling strength seems to vary widely, with some experiments showing only weak, if any, vibronic effects \cite{Ilani} while others exhibit well-pronounced vibrational sidebands or even the Franck-Condon blockade.\cite{Sapmaz,Huttel1,Huttel2,Leturcq}

Most of the corresponding theoretical analysis so far \cite{Sapmaz,Sapmaz2,Grifoni,FlensbergCNT} focused on an electron-vibron coupling of an {\em extrinsic} electrostatic origin. The basic idea behind this extrinsic coupling is the mechanical deformation of the nanotube induced by the electrostatic interaction between the charge on the nanotube and, e.g., a nearby gate electrode. These studies leave several issues unresolved which we address in this paper: (i) The estimates do not succeed in justifying the large values of the coupling constant $\lambda$ exhibited by some samples. (ii) A criterion to determine which vibron modes contribute mostly is still missing. (iii) A viable theory must explain the strong sample dependence of the electron-vibron coupling.
 
In this paper we address these issues by analyzing the contribution to the electron-vibron coupling originating from the {\em intrinsic} electron-phonon coupling in CNT. This intrinsic coupling originates from two principal mechanisms:\cite{Mahan,Suzuura,Harrison} (i) The deformation-induced modification of bond lengths induces modulations of the electron hopping in the honeycomb lattice. (ii) Distortions induce a local variation of areas and generate a deformation potential. It is interesting to mention that these two coupling mechanisms manifest themselves quite differently in an effective Dirac equation of the electronic degrees of freedom which we employ in this work. While the deformation potential enters the Dirac equation as a scalar potential, the bond length modifications enter as a gauge field.\cite{Mahan,Suzuura,Kane} 

The coupling constant of the deformation potential is estimated to be about an order of magnitude larger than the corresponding one for hopping modulations.\cite{Suzuura} When combined with symmetry arguments, this allows one to identify which modes couple most strongly to electrons. Longitudinal stretching modes as well as radial breathing modes involve local area variations and couple strongly via the deformation potential, while transverse twist modes involve pure shear and thus couple only via the weaker hopping-modulation mechanism. Finally, bending modes, though coupling via the dominant deformation potential, interact only quadratically with electrons, since for symmetry reasons their effect cannot depend on the sign of the associated deformation. Thus, longitudinal stretching modes and radial breathing modes are the natural candidates to couple strongly to electrons, in agreement with experimental conclusions.\cite{LeRoy,Sapmaz,Huttel1,Huttel2,Leturcq}

To quantify the strength of the electron-vibron coupling in suspended CNT quantum dots, we employ a mapping to an Anderson-Holstein model \cite{Boese,Mitra,Flensberg,Koch,KochLong} which appropriately describes transport through vibrating quantum dots with a discrete electronic spectrum. This model assumes that transport occurs through an isolated electronic state coupled to few vibrational modes of the nanostructure. 
The corresponding dimensionless coupling constant $\lambda$ is given by the shift of the vibron coordinate (induced by the tunneling electron) measured in units of the amplitude of its quantum fluctuations (i.e. the oscillator length).
An alternative parameter, often employed in experimental works is $g=\lambda^{2}_{}$. To lowest order, vibron modes will induce vibrational sidebands when the corresponding electron-vibron coupling $g$ is of order unity. The sidebands will be accompanied by the Franck-Condon blockade \cite{Koch,KochLong} --- a low bias suppression of the current --- when $g>1$.

The coupling parameter $g$ does not only depend on the specific vibron mode but is also sample specific, for at least two reasons: (i) It is sensitive to the geometrical details (radius or length) of the CNT, e.g., through the oscillator length of the vibron. (ii) It is sensitive to the location of the CNT quantum dot due to the strain profile of the vibron mode. Despite this sample dependence, we can give estimates of typical values of $g$ due to the intrinsic electron-vibron coupling for the different vibron modes of the CNT quantum dots. For the longitudinal stretching mode, we find upper limit estimates which are significantly larger than unity. This is consistent with the observation of vibrational sidebands \cite{Sapmaz,Huttel1,Huttel2,Leturcq}
and of the Franck-Condon blockade \cite{Leturcq}.  

The paper is organized as follows. In Sec.\ \ref{sec:review} we briefly review several ingredients required to analyze the electron-vibron coupling in carbon nanotube quantum dots. Specifically, this includes the elastic theory of long-wavelength phonons in CNT, the electronic properties of CNT in terms of the effective Dirac theory of graphene, the formation of CNT quantum dots, and the microscopic electron-phonon coupling in CNT. Sec.\ \ref{sec:elph} contains a detailed discussion of the dimensionless electron-vibron coupling constants of the various vibron modes for CNT quantum dots, and contains the central results of the paper. This section also discusses our results in relation to recent experiments. We conclude and summarize in Sec.\ \ref{sec:conclusions}. Some calculational details are relegated to an appendix.

\section{Phonons and electrons in CNT}
\label{sec:review}

\subsection{Vibron modes in suspended CNT}

\begin{figure}[ht]
	\centering
		\includegraphics[width=0.7\columnwidth]{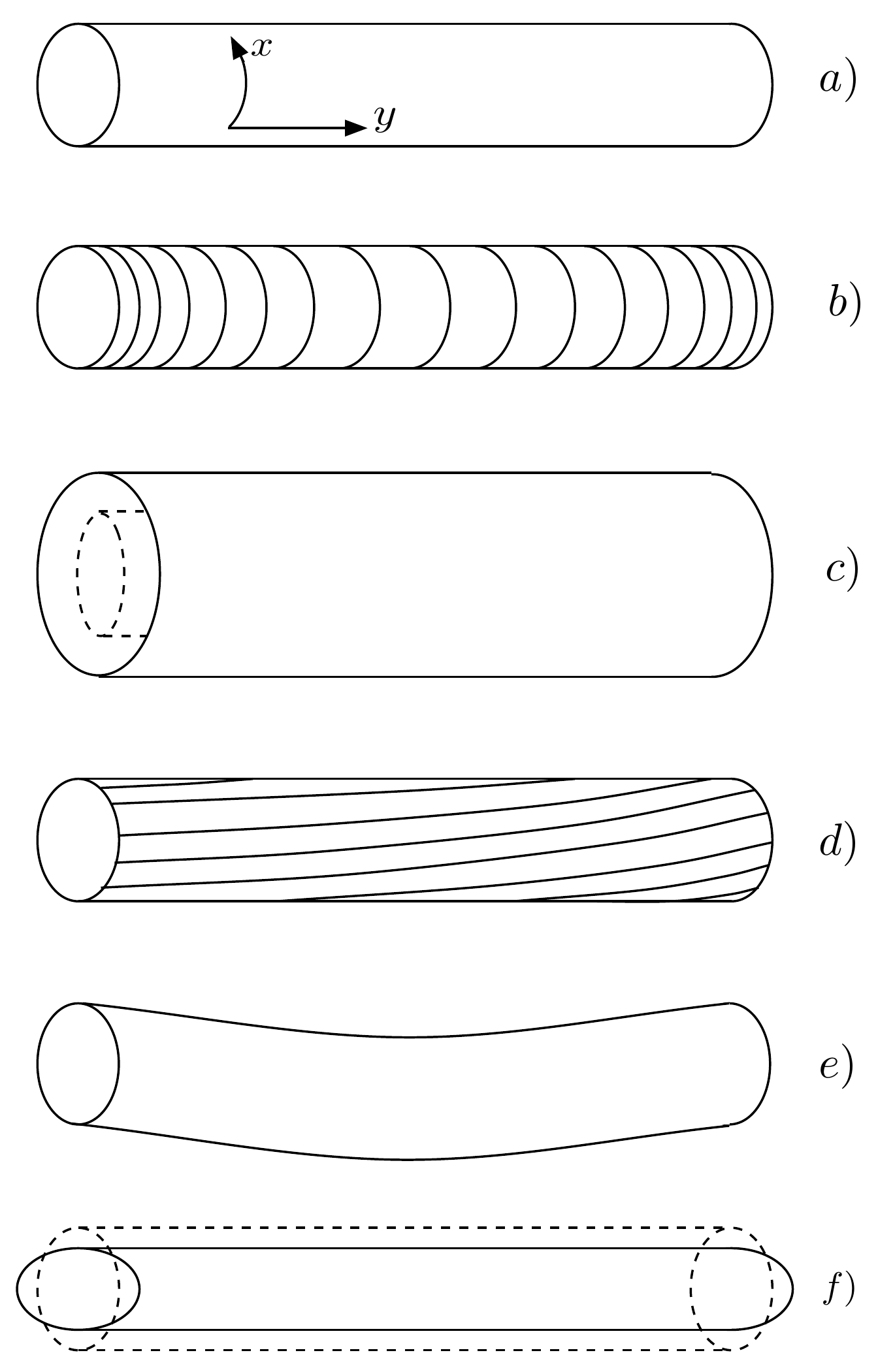}
			\caption{Schematic description of the acoustic phonon modes of CNT. a) Unperturbed nanotube. b) Stretching mode. c) Breathing mode. d) Twist mode. e) Bending mode with $n=1$ at finite $q$. f) Bending mode with $n=2$ and $q=0$.
			\label{Phonons}}
\end{figure}

The acoustic phonon branches of CNT can be described within the elastic theory of a cylindrical membrane. A CNT with radius $r$ can be viewed as a folded graphene
sheet described by the elastic Lagrangian density \cite{LandauBook,Mahan,Suzuura}
\begin{equation}
\label{elastic}
    {\cal L} = T-V_{{\rm stretch}}^{}-V_{{\rm bend}}^{}
\end{equation}
with
\begin{eqnarray}
\label{StretchBend}
&& T = \frac{\rho_0}{2} \left(\dot{\bf u}^2+\dot h^2\right)\nonumber \\
&& V_{{\rm stretch}}^{} = \mu u^2_{ij}+\frac{1}{2} \lambda u^2_{kk} \\
&& V_{{\rm bend}}^{} = \frac{1}{2}\kappa \left(\nabla^2 h+\frac{h}{r^{2}_{}}\right)^2\nonumber
\end{eqnarray}
in terms of the mass density $\rho_0$, the Lam\'e coefficients $\mu$ and $\lambda$ characterizing the
in-plane rigidity of the lattice, and the bending rigidity $\kappa$. In Eq.\ (\ref{StretchBend}), $u_{ij}^{}$ denotes the strain tensor of a cylindrical membrane whose components are \cite{LandauBook} 
\begin{eqnarray}
&& u_{xx}=\partial_{x}^{}u_{x}^{}+\frac{h}{r}\, ,\quad\quad
u_{yy}=\partial_{y}^{}u_{y}^{}\nonumber\\
&&\quad\quad\;\; 2\,
u_{xy}=\partial_{x}^{}u_{y}^{}+\partial_{y}^{}u_{x}^{}
\label{StrainTube}
\end{eqnarray}
up to linear order in the tangential and radial distortions $u_{i}^{}({\bf r})$ and $h({\bf r})$, respectively. Here we introduced $x$ and $y$ coordinates for the cylinder, fixing the $y$ axis along the CNT axis, while the $x$ axis is the curvilinear coordinate around its waist (see Fig.\ \ref{Phonons}). The strain tensor of a cylindrical membrane contains the additional term $h/r$ in $u_{xx}^{}$  with respect to a flat 2d membrane, describing the strain induced by a uniform change in $h$. In addition, the finite curvature of the CNT induces the extra term $h/r^{2}_{}$ in the bending energy. One readily finds that the strain vanishes for distortions described by
\begin{eqnarray}
\label{Shift}
&& u_{x}^{}=-\frac{u}{n}\,\sin \left( \frac{nx}{r}\right) \nonumber\\
&& u_{y}^{}={\rm constant}\\
&& h=u\,\cos \left( \frac{nx}{r}\right)\nonumber
\end{eqnarray}
with $n$ integer. In particular, the $n=1$ case describes a rigid translation of the tube in a direction perpendicular to its axis. The term $V_{{\rm bend}}^{}$ in Eq.\ (\ref{elastic}) describes the energy cost to induce a local {\em variation of curvature} with respect to the cylindrical geometry. This bending energy vanishes for the rigid shift Eq.\ (\ref{Shift}) with $n=1$, but produces a gap for the $n\ge 2$ modes which guarantees the stability of the cylindrical shape of the tube.

The Euler-Lagrange equations for the Lagrangian Eq.\ (\ref{elastic}) and  the resulting phononic branches are discussed in the Appendix. The distortions involved in stretching, breathing, twist and bending modes are depicted in Fig.\ (\ref{Phonons}).
The corresponding long-wavelength (i.e. $qr\ll 1$) dispersions are given by
\begin{eqnarray}
&&\omega_{n=0,qr\ll 1}^{({\rm Stretch})}=v_{{\rm Stretch}}^{}\, q\nonumber\\
&&\omega_{n=0,qr\ll 1}^{({\rm Twist})}=v_{{\rm Twist}}^{}\, q\nonumber\\
&&\omega_{n=0,qr\ll 1}^{({\rm Breath})}\simeq\omega_{{\rm B}}^{}=\sqrt{(Ar^{2}_{}+\kappa)/(\rho_{0}^{}r^{4}_{})}\nonumber\\
&&\omega_{n= 1,qr\ll 1}^{({\rm Bend})}\simeq \left(qr\right)^{2}_{}\sqrt{\frac{2\mu(\mu+\lambda)}{(2\mu +\lambda)\rho_{0}^{}r^{2}_{}}}\nonumber\\
&&\omega_{n\geq 2,qr\ll 1}^{({\rm Bend})}\simeq \left(\frac{n_{}^{2}-1}{r^{2}_{}}\right)\sqrt{\frac{\kappa}{\rho_{0}^{}}\cdot\frac{n_{}^{2}}{n^{2}_{}+1}}\; .
\label{ModesRescaled}
\end{eqnarray}
Here, the group velocities of the stretching and twist modes are $v_{{\rm Stretch}}^{}=\sqrt{(A^{2}_{}-\lambda^{2}_{})/(A\rho_{0}^{})}$ and $v_{{\rm Twist}}^{}=\sqrt{\mu /\rho_{0}^{}}$, with $A=2\mu +\lambda$. The $n=1$ bending mode has a quadratic dispersion reminiscent of the flexural phonon modes in graphene.\cite{Mariani} Unlike the graphene case, however, its dispersion is mostly determined by stretching energies due to the finite curvature of the nanotube. 
The dispersion of the breathing mode is characterized by a finite frequency $\omega_{\rm B}$ at long wavelengths.
For typical parameters,\cite{Suzuura,Parameters} one finds $Ar_{}^{2}\simeq 50\, {\rm eV}\cdot L_{\perp}^{2}[\rm{nm}]\gg\kappa$ for any realistic nanotube. The bending modulus can thus be neglected in the dispersion of the breathing mode. The corresponding $q=0$ energy gap is then $\hbar\omega_{{\rm B}}^{}
\simeq 8\cdot 10_{}^{-2}\,{\rm eV}/L_{\perp}^{}[\rm{nm}]$, with $L_{\perp}^{}=2\pi r$ denoting the circumference of the tube. As an example, the (10,10) armchair CNT with $r\simeq 7{\rm \AA}$ has $\hbar\omega_{{\rm B}}^{}\simeq 2\cdot 10_{}^{-2}\,{\rm eV}$, which makes the breathing mode frequently too high in energy to be excited in low-bias transport measurements.

In experiments, CNT are clamped by contacts to leads at two points separated by a distance $L$. The wavenumber quantization $q=m^{\prime}_{}\pi /L$ (with $m^{\prime}_{}$ an integer number) yields discrete phonon oscillator modes, called vibrons. These contacts will also frequently induce tension in the CNT which can be included into the elastic Lagrangian Eq.\ (\ref{StretchBend}) by the modification 
\begin{equation}
\label{Tension}
V_{{\rm bend}}^{}\rightarrow V_{{\rm bend}}^{}+\gamma \left(\partial_{y}^{}h\right)^{2}_{}\; ,
\end{equation}
where $\gamma$ measures the tension applied at the ends of the tube. The parameter $\gamma$ is thus sample as well as temperature dependent. Tension modifies the dispersion of the bending modes into
\begin{equation}
\label{BendingTension}
\omega_{n\ge 1,q}^{({\rm Bend,\, Tens})}=\sqrt{\omega_{n\ge 1,q}^{({\rm Bend})\, 2}+\frac{\gamma}{\rho_{0}^{}}q^{2}_{}}\, .
\end{equation}
The dispersion of the $n=1$ mode thus becomes linear at small momenta and crosses over to a quadratic dependence at higher momenta. The gapped bending modes with $n\ge 2$ are less sensitive to tension. 

\subsection{Electronic states in CNT quantum dots}

{\it Dirac Hamiltonian for CNT}.--- The electronic properties of CNT can be conveniently described in terms of the low energy effective Dirac Hamiltonian of electrons in graphene \cite{DiVincenzo}
\begin{equation}
\label{HDirac}
H=\hbar v\,\mathbf{\Sigma}\cdot\mathbf{k}
\end{equation}
with velocity $v\simeq 10^{6}_{}\,{\rm m\cdot s_{}^{-1}}$, where the 2d wavenumber $\mathbf{k}$ is measured from the relevant Dirac point $\pm {\bf k}_{D}^{}$, with $\mathbf{k}_{D}^{}=2\pi/(3\sqrt{3}a)\, (\sqrt{3},1)$ and $a=1.42\, \AA$ the bond length. The Hamiltonian in Eq.\ (\ref{HDirac}) acts on four-component spinors $(u_{A,\mathbf{k}}^{+},u_{B,\mathbf{k}}^{+},u_{B,\mathbf{k}}^{-},u_{A,\mathbf{k}}^{-})$ of Bloch amplitudes in the space spanned by the honeycomb sublattices ($A/B$) and Dirac valleys ($+/-$). The matrices $\Sigma_{x,y}^{}=\Pi_{z}^{}\otimes\mathbf{\sigma}_{x,y}^{}$ denote components of a vector $\mathbf{\Sigma}$, with $\Pi_{i}^{}$ and $\sigma_{j}^{}$ the Pauli matrices in the spaces of the Dirac valleys and the sublattices, respectively. In CNT the folding of the graphene sheet amounts to identifying different lattice sites of the same type ($A$ or $B$) separated by a lattice vector of the form ${\bf T}=\bar{m}{\bf T}_{1}^{}+\bar{n}{\bf T}_{2}^{}$, with ${\bf T}_{1}^{}=a\sqrt{3}\, (0,1)$ and ${\bf T}_{2}^{}=a\sqrt{3}/2\, (\sqrt{3},1)$ denoting two independent generators of the honeycomb lattice.\cite{Kane} The corresponding nanotube is denoted as $[\bar{m},\bar{n}]$ and it is customary to introduce the chiral angle $\theta =\arctan (T_{y}^{}/T_{x}^{})$. Thus armchair CNT correspond to $\theta =0$ while zig-zag ones stem from $\theta =\pi /2$. Owing to the underlying Bloch wavefunction, a translation by $\bf T$ modifies the Dirac spinors in the two valleys by a multiplicative factor $\exp [\pm i{\bf k}_{D}^{}\cdot {\bf T}]$. This fact can be accomodated in the Dirac description of CNT by formally introducing an Aharonov-Bohm vector potential ${\bf a}_{\pm}^{}$, which induces a fictitious magnetic flux $\phi$ given by $\Phi\equiv\phi /\phi_{0}^{}=\pm\mathbf{k}_{D}^{}\cdot {\bf T}/2\pi$.\cite{Kane} Here $\phi =\oint {\rm d}{\bf l}\cdot {\bf a}_{\pm}^{}$ with the line integral taken along a closed loop around the cylinder waist and $\phi_{0}^{}=h/e$ is the flux quantum. Due to flux periodicity, we can restrict attention to $\Phi\in [-1/2,1/2]$. Choosing the $x$-axis for the CNT along its waist, i.e. $\hat{{\bf x}}=\hat{{\bf T}}$, the electronic eigenstates are proportional to $\exp [i(2\pi lx/L_{\perp}^{} +ik_{y}^{}y)]$, where $l$ is an integer number. The low energy Hamiltonian for the CNT thus reads
\begin{equation}
\label{HDiracTube}
H=\Pi_{z}^{}\otimes \hbar v\,\left[\sigma_{x}^{}\frac{2\pi}{L_{\perp}^{}}(l+\Phi) +\sigma_{y}^{}k_{y}^{}\right]\; .
\end{equation}
By direct calculation for a $[\bar{m},\bar{n}]$ nanotube, one obtains $\Phi =\pm (\bar{m}+2\bar{n})/3$. When $\Phi$ is integer, the corresponding CNT are {\it metallic} characterized by a gapless and linear dispersion. Other values of $\bar{m}$ and $\bar{n}$ lead to {\it semiconducting CNT} with gapped electronic spectra, the smallest gap being given by $\Delta =\hbar v \Phi 2\pi/L_{\perp}^{}$. From now on we will assume that only the lowest energy subband $l=0$ is relevant. 

\begin{figure}[ht]
	\centering
		\includegraphics[width=0.9\columnwidth]{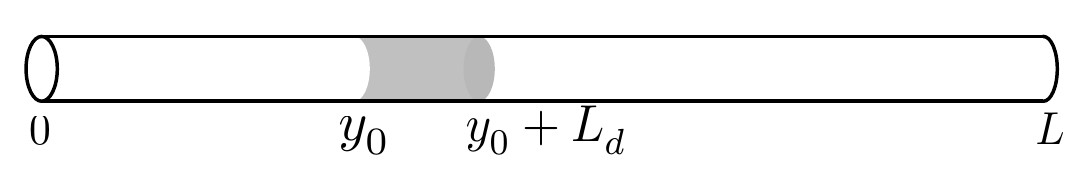}
			\caption{Schematic description of the electronic quantum dot of size $L_{d}^{}$ (in gray) along a carbon nanotube of length $L$.
			\label{CNTdot}}
\end{figure}

{\it CNT quantum dots}.---The possibility of opening a gap in CNT allows for the formation of CNT quantum dots. 
This can be achieved by electrostatic  potentials for semiconducting CNT or by inhomogeneities in the graphene folding for metallic CNT. In general, the location of the quantum dots along the CNT and their length $L_{d}^{}$ have no relation with the overall nanotube size $L$ and can thus be sample specific, as indicated in Fig. \ref{CNTdot}. 

In {\it semiconducting CNT} electron confinement can be induced by an electrostatic potential $V(y)$ (induced intentionally by gates or by disorder) along the tube direction which enters the Dirac Hamiltonian as
\begin{equation}
\label{NanoDot}
H=\left(\begin{array}{cc}V(y) & \Delta-i\hbar v k_{y}^{} \\\Delta+i\hbar v k_{y}^{} & V(y)\end{array}\right)\; .
\end{equation}
Here we consider potentials smooth on the scale of the honeycomb lattice spacing which allows us to restrict attention to a single cone of the Dirac description. The physics of electron confinement can be captured by the simplest choice for $V(y)$, namely a square well potential of magnitude $V(y)=-V_{0}^{}$ between $y=y_{0}^{}$ and $y=y_{0}^{}+L_{d}^{}$ and zero elsewhere, with $y_{0}^{}$ the location of the left edge of the electronic dot and $L_{d}^{}$ its extension along the tube direction. (Of course, within the single Dirac cone approximation the abrupt variation of the potential still needs to be slow on the scale of the bond length $a$.) A potential with $V_{0}^{}>\Delta /2$ confines electrons in the dot and depletes occupation outside the region $y\in [y_{0}^{},y_{0}^{}+L_{d}^{}]$, yielding quantization of the longitudinal wavenumber $k_{y}^{}=n^{\prime}_{}\pi /L_{d}^{}$, with $n^{\prime}_{}\ge 1$ an integer. In the limit $\hbar v k_{y}^{}\ll \Delta$ one gets a vanishing wavefunction for $y<y_{0}^{}$ and $y>y_{0}^{}+L_{d}^{}$ and a non-vanishing spinor $|n^{\prime}_{}\rangle$ for $y\in [y_{0}^{},y_{0}^{}+L_{d}^{}]$
\begin{equation}
\label{SpinorDot}
\psi_{n^{\prime}_{}}^{}(y)=\langle y|n^{\prime}_{}\rangle=\frac{1}{\sqrt{L_{d}^{}}}\left(\begin{array}{c}1 \\1\end{array}\right)\sin \left(\frac{n^{\prime}_{}\pi (y-y_{0}^{})}{L_{d}^{}}\right) 
\end{equation}
with energy
\begin{equation}
\label{EnerDot}
E_{n^{\prime}_{}}^{}=\Delta-V_{0}^{}+\frac{\hbar^{2}_{} v^{2}_{} n^{\prime 2}_{}\pi^{2}_{}}{2L_{d}^{2}\Delta}\; .
\end{equation}

\begin{figure}[ht]
	\centering
		\includegraphics[width=0.7\columnwidth]{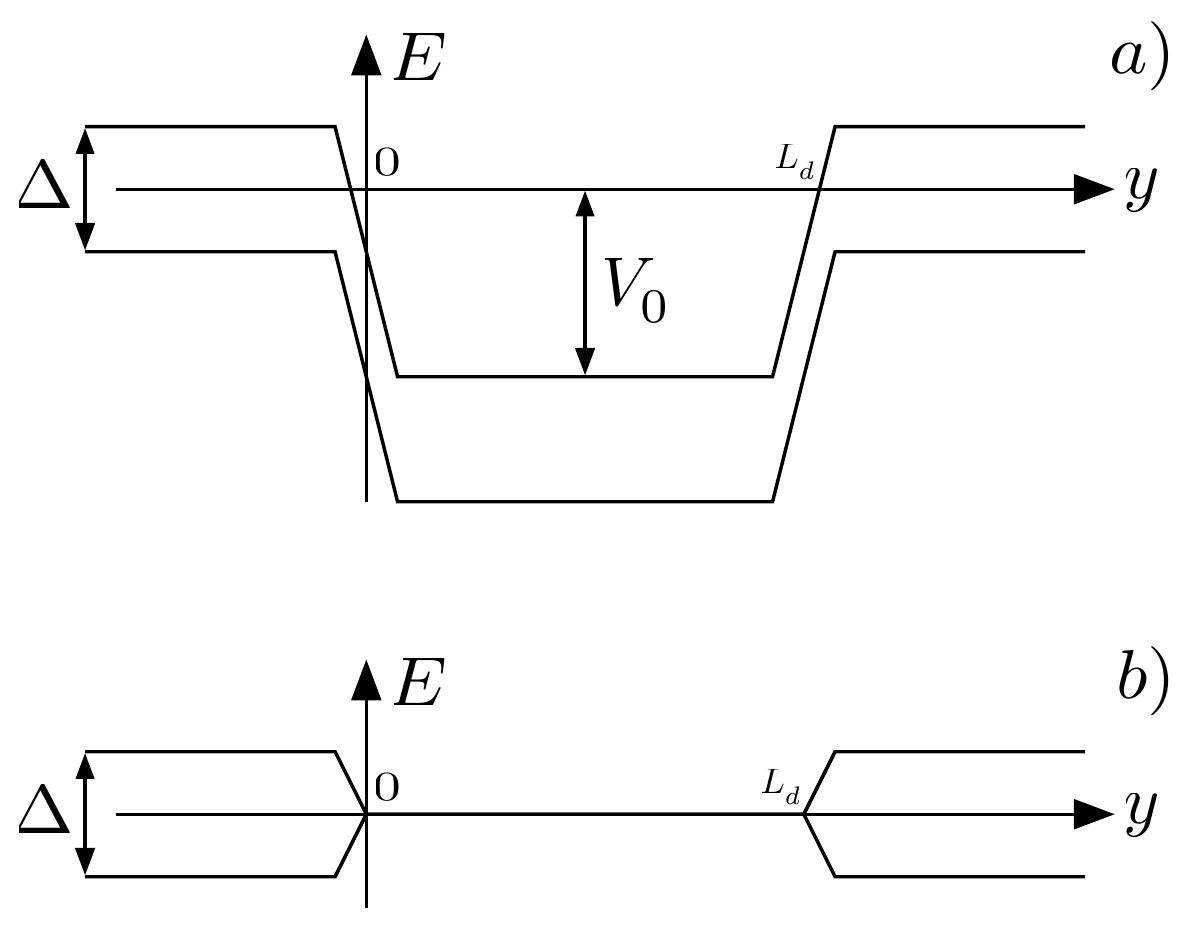}
			\caption{Schematic description of the electron confinement producing quantum dots in semiconducting (a) and metallic (b) carbon nanotubes. Here $y_{0}^{}=0$.
			\label{GapDot}}
\end{figure}

In {\it metallic CNT}, quantum dots can be induced by a non-uniformity in the folding vector ${\bf T}$. If $\Phi$ is non-integer for $y<y_{0}^{}$ and $y>y_{0}^{}+L_{d}^{}$, and integer for $y\in [y_{0}^{},y_{0}^{}+L_{d}^{}]$, the gap opening outside the dot confines electrons within the metallic region with integer $\Phi$. The wavefunction for $y\in [y_{0}^{},y_{0}^{}+L_{d}^{}]$ is then
\begin{equation}
\label{SpinorDotMet}
\psi_{n^{\prime}_{}}^{}(y)=\langle y|n^{\prime}_{}\rangle=\frac{1}{\sqrt{L_{d}^{}}}\left(\begin{array}{c}1 \\i\end{array}\right)\sin \left(\frac{n^{\prime}_{}\pi (y-y_{0}^{})}{L_{d}^{}}\right) 
\end{equation}
in the limit $\hbar v k_{y}^{}\ll \Delta$, with a corresponding energy
\begin{equation}
\label{EnerDotMet}
E_{n^{\prime}_{}}^{}=\hbar v \frac{n^{\prime}_{}\pi}{L_{d}^{}}\quad .
\end{equation}
The potential distribution and energy profiles involved in the creation of quantum dots in CNT are sketched in Fig.\ \ref{GapDot}.
While vibron modes of the nanotube involve its entire length, the location and extension of the electronic quantum dot are essentially determined by inhomogeneities and are in principle uncorrelated with the nanotube size. This effect is particularly pronounced in dirty CNT. 

\subsection{Electron-phonon coupling in CNT}

The coupling between electrons and phonons in graphene and in nanotubes
can be induced by two different mechanisms.\cite{Mahan,Suzuura,Mariani}

(i) The deformation potential \cite{Harrison} is induced by the local area variation of the membrane which is associated with a phonon excitation. Within the Dirac representation in a single valley, the deformation potential can be written as \cite{Mahan,Suzuura,Mariani}
\begin{equation}
V^{({\rm def})}_{e-ph}=g_{1}^{}\left(u_{xx}^{}+u_{yy}^{}\right)\,
{\bf 1}\quad .
\end{equation}
The corresponding coupling strength has been estimated to be as large as $g_{1}^{}\simeq 30\, {\rm eV}$.\cite{Suzuura}

(ii) Phonon modes involving pure shear do not induce a local variation of area and the associated strain has a vanishing trace. These modes couple to electrons only via the distortion-induced modification of bond lengths in the honeycomb lattice. The corresponding modulation in the electron hopping translates in the Dirac language into a purely off-diagonal term, correcting the electronic momentum in the same way as a (fictitious) gauge field, with vector potential \cite{Mahan,Suzuura,Kane,Mariani}
\begin{eqnarray}
  e{\bf A}({\bf r},t)&=& \frac{1}{2}\frac{\hbar}{t}\frac
{\partial t}{\partial a}D[-3\theta]\left[\begin{array}{c} 2 u_{xy} \\
  (u_{xx}-u_{yy})
  \end{array}\right]\; .
\label{fictitious}
\end{eqnarray}
Here $t$ denotes the hopping amplitude, while the rotation matrix $D[-3\theta]=\cos 3\theta\,{\bf 1}+i\sin 3\theta\, \sigma_{y}^{}$ is due to the rotation of the folding vector ${\bf T}$ by the angle $\theta$ with respect to the $x-y$ axes in graphene. The form of the vector potential is essentially fixed by the honeycomb lattice symmetry. Combined with the diagonal deformation potential, the fictitious gauge field yields the total electron-phonon coupling matrix
\begin{equation}
 \label{Velph}
  V_{e-ph}^{}=  \left(\begin{array}{cc}g_{1}^{}(u_{xx}+u_{yy}) & g_
{2}^{}e^{i3\theta}_{}f^{*}_{}[u_{ij}^{}]
\\g_{2}^{}e^{-i3\theta}_{}f[u_{ij}^{}] & g_{1}^{}(u_{xx}+u_{yy})\end{array}\right)\; ,
\end{equation}
with $f[u_{ij}^{}]=2u_{xy}+i (u_{xx}-u_{yy})$
and the hopping-modulation coupling constant $g_{2}^{}=\frac{\hbar v}{2t}\frac{\partial t}{\partial a}\simeq 1.5\, {\rm eV}$.\cite{Suzuura} The latter is estimated to be much smaller than the corresponding one for the deformation potential. 

Thus, {\it stretching modes} and {\it breathing modes} induce local variations of areas, yielding a strong coupling via the deformation  potential, while {\it twist modes} involve pure shear and couple to electrons only via the weaker "gauge-field coupling". For reasons of symmetry, {\it bending modes} are qualitatively different, as their effect  on the electronic energy {\it cannot} depend on the sign of the associated deformation. Indeed, we will show that they yield a vanishing expectation value of the strain on the electronic states in first order in the displacement fields. Hence, bending modes require an analysis of the strain tensor of the cylindrical membrane to quadratic order. A lengthy but straightforward calculation yields
\begin{eqnarray}\label{StrainSecond}
u_{xx}^{}&=&\partial_{x}^{}u_{x}^{}+\frac{h}{r}+\nonumber\\
&&+\frac{1}{2}\left[ (\partial_{x}^{}h-\frac{u_{x}^{}}{r})^{2}+(\frac{h_{}^{}}{r_{}^{}}+\partial_{x}^{}u_{x}^{})^{2}_{}+(\partial_{x}^{}u_{y}^{})^{2}_{}\right]\nonumber\\
u_{yy}^{}&=&\partial_{y}^{}u_{y}^{}+\frac{1}{2}\left[ (\partial_{y}^{}h)^{2}_{}+(\partial_{y}^{}u_{x}^{})^{2}_{}+(\partial_{y}^{}u_{y}^{})^{2}_{}\right]\nonumber\\
2u_{xy}^{}&=&\partial_{x}^{}u_{y}^{}+\partial_{y}^{}u_{x}^{}+\nonumber\\
&&+(\partial_{x}^{}h)(\partial_{y}^{}h)-\frac{u_{x}^{}}{r}(\partial_{y}^{}h)+\frac{h}{r}(\partial_{y}^{}u_{x}^{})+\nonumber\\
&&+(\partial_{x}^{}u_{x}^{})(\partial_{y}^{}u_{x}^{})+(\partial_{x}^{}u_{y}^{})(\partial_{y}^{}u_{y}^{})\; .\nonumber
\end{eqnarray}
One readily checks that this result respects the requirement that the rigid translation described by Eq.\ (\ref{Shift}) with $n=1$ is strain-free. As will be discussed in detail in the next session, for the $n=1$ bending mode a finite coupling mediated predominantly by the deformation potential can be obtained only at finite $q$, while modes with $n\ge 2$ couple also for $q=0$. Having established the form of the microscopic electron-phonon interaction in the language of the Dirac hamiltonian, we are now ready to systematically analyze the strength of the coupling for different types of vibrons in CNT quantum dots.

\section{Electron-vibron coupling in CNT quantum dots}
\label{sec:elph}

Electronic transport through quantum dots located in suspended carbon nanotubes is adequately described by an (extended) Anderson-Holstein model. Motivated by experiments, we will focus on the limit in which the vibrational frequencies are small compared to the electronic level spacing.\cite{footnote} Then, we can effectively restrict attention to a single electronic state coupled to several vibrational modes. We emphasize, however, that the parameters of the model can depend on the particular electronic state under consideration. 

We now use the results reviewed in the previous section in order to evaluate the electron-vibron coupling entering into the Anderson-Holstein Hamiltonian for a CNT quantum dot \cite{Mitra,Flensberg,Koch,KochLong}
\begin{eqnarray}
\label{Anderson}
&&H=H_{{\rm el}}^{}+H_{{\rm vib}}^{}+H_{{\rm leads}}^{}+H_{{\rm T}}^{}\nonumber\\
&&H_{{\rm el}}^{}=\sum_{s=\uparrow ,\downarrow }^{}\sum_{\nu , q}^{}\left[\epsilon +\lambda^{(\nu )}_{q}\hbar\omega^{(\nu )}_{q} (b_{q}^{(\nu)\dagger}+b_{q}^{(\nu )})^{j}_{}\right] n^{}_{s}+Un_{\uparrow}^{}n_{\downarrow}^{}\nonumber\\
&&H_{{\rm vib}}^{}=\sum_{\nu , q}^{}\hbar\omega^{(\nu )}_{q}\, b_{q}^{(\nu)\dagger}b_{q}^{(\nu )}\; .
\end{eqnarray}
Here $H_{{\rm el}}^{}+H_{{\rm vib}}^{}$ describes the electronic and vibronic properties of the CNT quantum dot, while $H_{{\rm leads}}^{}+H_{{\rm T}}^{}$ describes the leads and the tunneling of electrons between the leads and the dot.
In Eq.\ (\ref{Anderson}), $n_{s}^{}$ is the electron number operator with spin $s$, $\epsilon$ the energy of the electron state, and $U$ the repulsion energy for double occupancy of the dot. The operator $b_{q}^{(\nu )}$ annihilates the vibron mode of type $\nu$ ($\nu$ = Stretch, Twist, Breath, Bend) and wavenumber $q$, with energy $\hbar\omega_{q}^{(\nu)}$. The electron-vibron interaction in $H_{{\rm el}}^{}$ can be linear ($j=1$) or quadratic ($j=2$) in the vibronic operators, depending on the vibron mode under consideration. 

In order to compute the dimensionless electron-vibron coupling constant $\lambda_{q}^{(\nu )}$, we rewrite the electronic Hamiltonian $H_{\rm el}$ in terms of the vibronic displacement operator $u_{q}^{(\nu )}=(b_{q}^{(\nu )}+b_{q}^{(\nu )\dagger})\, l_{{\rm osc},q}^{(\nu )}/\sqrt{2}$ as
\begin{equation}
\label{AndElVib}
H_{{\rm el}}=\sum_{s=\uparrow ,\downarrow }^{}\sum_{\nu , q}^{}[\epsilon+\lambda^{(\nu )}_{q}\left( \frac{\sqrt{2}{u_{q}^{(\nu )}}}{l_{{\rm osc}, q}^{(\nu )}}\right)^{j}_{}\hbar\omega^{(\nu )}_{q}] \, n_{s}^{}+Un_{\uparrow}^{}n_{\downarrow}^{}\quad , 
\end{equation}
with the oscillator length $l_{{\rm osc},q}^{(\nu )}=\sqrt{\hbar /M\omega_{q}^{(\nu )}}$ and $M=\rho_{0}^{}LL_{\perp}^{}$ the oscillator mass. We can now identify the electron-vibron coupling constant $\lambda^{(\nu )}_{q}$ with the shift in the energy of the localized electronic state induced by a static vibron displacement $u^{(\nu )}_{q}=l_{{\rm osc}, q}^{(\nu )}/\sqrt{2}$, measured in units of $\hbar\omega^{(\nu )}_{q}$. This shift can be readily computed in perturbation theory, employing the intrinsic electron-vibron coupling $V_{e-ph}^{}$ in Eq.\ (\ref{Velph}). We find that first-order perturbation theory suffices. Thus, an electronic state $|n'\rangle$ of the dot is shifted by $\Delta \epsilon_{q, n^{\prime}_{}}^{} (u^{(\nu )}_{q})=\langle n^{\prime}_{}|V_{e-ph}^{}(u^{(\nu )}_{q})|n^{\prime}_{}\rangle$ for a given vibron displacement $u_{q}^{(\nu )}$. E.g., for a semiconducting dot, the expectation value is taken with the electronic state Eq.\ (\ref{SpinorDot}). As a result, we obtain the expression 
\begin{equation}
\label{Lambda}
\lambda^{(\nu )}_{q, n^{\prime}_{}} =\frac{\Delta \epsilon_{q, n^{\prime}_{}}^{}(u^{(\nu )}_{q}=l_{{\rm osc},q}^{(\nu )}/\sqrt{2})}{\hbar\omega^{(\nu )}_{q}}
\end{equation}
for the dimensionless coupling constant $\lambda^{(\nu )}_{q, n^{\prime}_{}}$. 

In order to analyze the value of $\lambda^{(\nu )}_{q, n^{\prime}_{}}$ for the various vibron modes, we assume that the nanotube extends between $y=0$ and $y=L$, while the electronic wavefunction in the dot exists in the interval $y\in [y_{0}^{},y_{0}^{}+L_{d}^{}]$, as depicted in Fig.\ (\ref{CNTdot}). The various vibrons have a quantized wavenumber $q=m^{\prime}_{}\pi/L$ along $y$, with $m^{\prime}_{}$ an integer number (the index $q$ will be replaced by the corresponding $m^{\prime}_{}$ from now on). We start considering the $n=0$ modes.

{\it Stretching mode}: The stretching mode
implies local area variations inducing a deformation potential together with hopping modulations which amount, in the Dirac language, to the electron-phonon coupling  matrix
\begin{equation}
  V_{e-ph}^{({\rm Stretch})}=\left(\begin{array}{cc}g_{1}^{}\partial_{y}^{}u_{y}^{} & ig_{2}^{}\partial_{y}^{}u_{y}^{}e^{i3\theta}_{}
\\ -ig_{2}^{}\partial_{y}^{}u_{y}^{}e^{-i3\theta}_{} & g_{1}^{}\partial_{y}^{}u_{y}^{}\end{array}\right)\; . \\
  \label{VelphStretch}
\end{equation}
Within the $qr\ll 1$ regime, the stretching mode is described by $u_{y}^{}=u\sin(m^{\prime}_{}\pi y/L)$, $u_{x}^{}=0$ and $h\simeq -i qr\lambda/A u_{y}^{}\ll u_{y}^{}$. Its linear coupling to electrons is dominated by the $u_{yy}^{}$ component in the deformation potential (while we neglect $u_{xx}^{}\simeq -\lambda/A u_{yy}^{}$ due to the small ratio $\lambda/A\simeq 0.1$), resulting in the electronic energy shift
\begin{equation}
\Delta \epsilon^{{\rm (Stretch)}}_{m^{\prime}_{}, n^{\prime}_{}}(u) =2g_{1}^{}u\frac{m^{\prime}_{}\pi}{L}\frac{I_{m^{\prime}_{},n^{\prime}_{}}^{}}{L_{d}^{}}
\end{equation}
where we introduced
\begin{eqnarray}
\label{Imn}
&&I_{m^{\prime}_{}, n^{\prime}_{}}^{}=\int_{y_{0}^{}}^{y_{0}^{}+L_{d}^{}}{\rm d}y\cos\left(\frac{m^{\prime}_{}\pi y}{L}\right)\sin^{2}_{}\left(\frac{n^{\prime}_{}\pi (y -y_{0}^{})}{L_{d}^{}}\right)=\nonumber\\
&&\quad=\frac{L}{m^{\prime}_{}\pi}\frac{\sin\left(\frac{m^{\prime}_{}\pi L_{d}^{}}{2L}\right)\cos\left(\frac{m^{\prime}_{}\pi \left( 2y_{0}^{}+L_{d}^{}\right)}{2L}\right)}{1-\left(\frac{m^{\prime}_{} L_{d}^{}}{2n^{\prime}_{}L}\right)^{2}_{}}\; .
\end{eqnarray}
Exploiting Eq.\ (\ref{Lambda}) we deduce 
\begin{equation}
\label{LambdaStretch}
\lambda_{m^{\prime}_{}, n^{\prime}_{}}^{({\rm Stretch})}= \frac{\sqrt{2}g_{1}^{}l_{{\rm osc}, m^{\prime}}^{({\rm Stretch})}}{\hbar v_{{\rm Stretch}}^{}}\frac{I_{m^{\prime}_{},n^{\prime}_{}}^{}}{L_{d}^{}}\; .
\end{equation}
Due to the coupling to electrons mediated by the dominant deformation potential and to their gapless dispersion, stretching vibrons are natural candidates to produce a large electron-vibron coupling. The result in Eq.\ (\ref{LambdaStretch}) shows that the coupling constant can be strongly sample specific, as it crucially depends on the geometric details of the nanotube and on the location and size of the CNT quantum dot (hidden in $I_{m^{\prime}_{},n^{\prime}_{}}^{}$).

In order to address this point we first consider the regime of very clean CNT, in which the dot size coincides with the total nanotube length (i.e. $y_0^{}=0$, $L_{d}^{}=L$). In this case we have $I_{m^{\prime}_{},n^{\prime}_{}}^{(L_{d}^{}=L,\, y_{0}^{}=0)}=-\delta_{m^{\prime}_{},2n^{\prime}_{}}^{}\, L/4$, yielding 
\begin{equation}
\label{LambdaStretchClean}
\lambda_{m^{\prime}_{}, n^{\prime}_{}, {\rm clean}}^{({\rm Stretch})}= -\delta_{m^{\prime}_{},2n^{\prime}_{}}^{}\frac{g_{1}^{}l_{{\rm osc}, m^{\prime}}^{({\rm Stretch})}}{2\sqrt{2}\hbar v_{{\rm Stretch}}^{}}\; .
\end{equation}
%
The interference between vibronic and electronic wavefunctions thus leads to strong selection rules suppressing the contribution of most vibron modes. The detectable ones are characterized by a coupling constant
\begin{equation}
\label{LambdaCleanStretchNum}
\lambda_{m^{\prime}_{},n^{\prime}_{}, {\rm clean}}^{({\rm Stretch})}\simeq -\delta_{m^{\prime}_{},2n^{\prime}_{}}^{}\frac{1.5}{\sqrt{m^{\prime}_{}L_{\perp}^{}[{\rm nm}]}}\; .
\end{equation}
Only the long wavelength modes in thin CNT can reach the strong coupling regime where vibron signatures in transport may emerge.

As the location and size of the quantum dot is not in general related to the overall CNT length, we can consider the opposite limit of a small dot, $L_{d}^{}\ll L$, where we obtain
\begin{equation}
\label{Imnlim}
I_{m^{\prime}_{},n^{\prime}_{}}^{(L_{d}^{}\ll L)}\simeq\frac{L_{d}^{}}{2}\cos\left(\frac{m^{\prime}_{}\pi y_{0}^{}}{L}\right)\; .
\end{equation}
This result stresses once more the strong sample-dependence of the electron-vibron coupling strength. If the dot localizes in regions involving weak vibron-induced strain the coupling constant $\lambda$ can be extremely small.
In contrast, dots localized around the regions of maximal strain in the tube can reach values up to $\Delta \epsilon^{{\rm (Stretch)}}_{m^{\prime}_{},n^{\prime}_{}, {\rm max}}(u)=g_{1}^{}u m^{\prime}_{}\pi /L$.
This leads to a maximal coupling 
\begin{equation}
\label{LambdaMaxStretchNum}
\lambda_{m^{\prime}_{},n^{\prime}_{}, {\rm max}}^{({\rm Stretch})}\simeq \frac{3}{\sqrt{m^{\prime}_{}L_{\perp}^{}[{\rm nm}]}}\; ,
\end{equation}
where we used the stretching mode dispersion (\ref{ModesRescaled}) in the oscillator length. We point out that vibron branches with a linear dispersion have a coupling which depends only on the nanotube circumference, not on its length. 
Remarkably, this coupling constant can be quite large for thin nanotubes with typical $L_{\perp}^{}$ in the nanometer range. For them, the measured coupling constant  $g=\lambda^{2}_{}$ for the lowest vibron mode can be of order 10, in agreement with recent experiments.\cite{Sapmaz,Huttel1,Huttel2,Leturcq} While we focus on the coupling of vibrons with localized electronic wavefunctions, the deformation potential contribution to the coupling between the stretching mode and the {\it bosonic excitations} of a Luttinger model for a very long CNT has been discussed previously.\cite{Grifoni} This analysis however resulted in a very small coupling constant.

{\it Breathing mode}: The breathing mode has a very weak momentum dependence in the $qr\ll 1$ regime. It is described by $u_{x}^{}=u_{y}^{}=0$, $h=u$ and by the coupling  
\begin{equation}
  V_{e-ph}^{({\rm Breath})}=\frac{h}{r}\left(\begin{array}{cc}g_{1}^{} & -i g_
{2}^{}e^{i3\theta}_{}
\\ i g_{2}^{}e^{-i3\theta}_{} & g_{1}^{}\end{array}\right)\; ,
  \label{VelphBreath}
\end{equation}
with a dominant deformation potential component. A similar analysis to that for the stretch mode above leads to $\Delta \epsilon^{{\rm (Breath)}}_{}(u)=g_{1}^{}u/r$ and to the coupling constant
\begin{equation}
\label{LambdaMaxBreathNum}
\lambda_{}^{({\rm Breath})}= \frac{g_{1}^{}l_{{\rm osc}}^{({\rm Breath})}}{\sqrt{2}r\hbar \omega_{{\rm B}}^{}}
\simeq\frac{7\cdot 10^{-2}_{}}{\sqrt{L[{\rm \mu m}]}}\; .
\end{equation}
Contrary to the stretching mode, this coupling constant depends only on the total tube length but is otherwise independent of the vibronic and electronic wavenumbers and is not affected by the localization properties of the electronic wavefunction. With this respect, the breathing mode dimensionless coupling is a much less sample specific quantity. Despite the coupling being mediated by the deformation potential, the large breathing mode gap significantly suppresses the oscillator length as well as the final value of the electron-vibron coupling constant.

{\it Twist mode}: Twist modes are associated with deformations given by $u_{x}^{}=u\sin(qy)$ and $u_{y}^{}=h=0$. Thus they induce a pure shear yielding, for a given displacement, a hopping-modulation coupling [see Eq.\ (\ref{Velph})]
\begin{equation}
  V_{e-ph}^{({\rm Twist})}=  g_{2}^{}\left(\begin{array}{cc}0 & e^{i3\theta}_{}\partial_{y}^{}u_{x}^{}
\\ e^{-i3\theta}_{}\partial_{y}^{}u_{x}^{} & 0\end{array}\right)
  \label{VelphTwist}
\end{equation}
to the electronic Dirac equation. Focusing on semiconducting dots, this implies \begin{equation}
\label{DeltaETwist}
\Delta \epsilon^{{\rm (Twist)}}_{m^{\prime}_{},n^{\prime}_{}}(u)=2g_{2}^{}u\frac{m^{\prime}_{}\pi}{L}\cos(3\theta)\frac{I_{m^{\prime}_{},n^{\prime}_{}}^{}}{L_{d}^{}}\; .
\end{equation}
As a result, using Eq.\ (\ref{Lambda}) we deduce 
\begin{equation}
\label{LambdaTwist}
\lambda_{m^{\prime}_{}, n^{\prime}_{}}^{({\rm Twist})}= \frac{\sqrt{2}g_{2}^{}l_{{\rm osc}, m^{\prime}}^{({\rm Twist})}\cos(3\theta)}{\hbar v_{{\rm Twist}}^{}}\frac{I_{m^{\prime}_{},n^{\prime}_{}}^{}}{L_{d}^{}}\; .
\end{equation}
Similarly to stretching modes, the linear dispersion of twist phonons yields a coupling constant dependent only on the nanotube circumference. However, the smallness of the fictitious gauge field coupling with respect to the deformation potential will in the end suppress the twist mode electron-vibron coupling. In parallel, the latter is sensitive to the CNT chirality via the $\cos(3\theta)$ factor, yielding a maximal effect on metallic armchair CNT (with $\theta =0$) and a vanishing one for zig-zag CNT (characterized by $\theta =\pi /2$). In the regime of a small dot localized around the maximum shear, using Eq.\ (\ref{Imnlim}) one can estimate the upper value of the dimensionless coupling
\begin{equation}
\label{LambdaMaxTwist}
\lambda^{{\rm (Twist)}}_{m^{\prime}_{},n^{\prime}_{}, {\rm max}}=\frac{g_{2}^{}l_{{\rm osc}, m^{\prime}_{}}^{{\rm (Twist)}}\cos(3\theta)}{\sqrt{2}\hbar v_{{\rm Twist}}^{}}\; .
\end{equation}
Replacing the typical parameters of graphene \cite{Parameters} in the oscillator length for the twist vibron we get
\begin{equation}
\label{LambdaMaxTwistNum}
\lambda_{m^{\prime}_{},n^{\prime}_{}, {\rm max}}^{({\rm Twist})}\simeq 0.17\frac{\cos(3\theta)}{\sqrt{m^{\prime}_{}L_{\perp}^{}[{\rm nm}]}}\; .
\end{equation}

{\it Bending modes}: The displacements involved in the bending vibrons with $n\ge 1$ are described in detail in the Appendix. It is crucial to notice that at finite momentum $q$ a non-vanishing component $u_{y}^{}$ is needed in order to correctly obtain the dispersion of the $n=1$ mode, while the gapped $n\ge 2$ branches are well described by $u_{x}^{}=-u/n\, \sin(nx/r)\sin(qy)$,  $h=u\cos(nx/r)\sin(qy)$ and $u_{y}^{}=0$ in the long wavelength regime $qr\ll 1$ for $\kappa\ll Ar^{2}_{}$. 
For $n=1$ the $u_{xx}^{}$ and $u_{yy}^{}$ components of the strain tensor up to linear order in the vibron displacement are proportional to $(qr)^{2}_{} \cos(x/r)\sin(qy)$, while $u_{xy}^{}\propto (qr)^{3}_{} \sin(x/r)\cos(qy)$. Thus they yield a vanishing expectation value on the $x$-independent electronic states.
In complete analogy, for $n\ge 2$ the expectation value of each component of the strain tensor on the electronic states vanishes in linear order in the amplitude of the distortions.  
We thus need to consider the strain tensor up to quadratic order as discussed in Eq.\ \ref{StrainSecond}. The expectation value of $u_{xy}^{}$ vanishes again, while those of $u_{xx}^{}$ and $u_{yy}^{}$ do not, giving rise to a non-vanishing deformation potential. From the $j=2$ case in Eq.\ (\ref{AndElVib}) we get the dominant energy corrections for $qr\ll 1$,
\begin{eqnarray}
&&\Delta \epsilon^{{\rm (Bend, n=1)}}_{m^{\prime}_{},n^{\prime}_{}}(u)=3g_{1}^{}\left(\frac{ u\pi m^{\prime }_{}}{2L}\right)^{2}_{}\left[\frac{1}{2}+\frac{I_{2m^{\prime}_{},n^{\prime}_{}}^{}}{L_{d}^{}}\right]\nonumber\\
&&\Delta \epsilon^{{\rm (Bend,n\ge 2)}}_{m^{\prime}_{},n^{\prime}_{}}(u)=g_{1}^{}\frac{u^{2}_{}}{4r^{2}_{}}\left(n^{}_{}-\frac{1}{n^{}_{}}\right)^{2}_{}\left[\frac{1}{2}-\frac{I_{2m^{\prime}_{},n^{\prime}_{}}^{}}{L_{d}^{}}\right]\nonumber\; .
\end{eqnarray}
The soft bending modes with $n=1$ behave differently from those with $n\ge 2$. The former have a gapless dispersion and a big oscillator length, while the latter are gapped and have much smaller quantum fluctuations. The resulting coupling constants are
\begin{eqnarray}
\label{LambdaBend}
&&\lambda_{m^{\prime}_{},n^{\prime}_{}}^{({\rm Bend}, n=1)}=g_{1}^{}\,\frac{3(2\mu +\lambda)}{4\mu (\mu +\lambda )}\,\frac{L}{m^{\prime 2}_{}L_{\perp}^{3}}\left[\frac{1}{2}+\frac{I_{2m^{\prime}_{},n^{\prime}_{}}^{}}{L_{d}^{}}\right]\nonumber\\
&&\lambda_{m^{\prime}_{},n^{\prime}_{}}^{({\rm Bend}, n\ge 2)}=\frac{g_{1}^{}}{\kappa}\,\frac{(n^{2}_{}+1)}{32\pi^{2}_{}n^{4}_{}}\,\frac{L_{\perp}^{}}{L}\left[\frac{1}{2}-\frac{I_{2m^{\prime}_{},n^{\prime}_{}}^{}}{L_{d}^{}}\right]\; .
\end{eqnarray}
It has to be pointed out that the functions $\frac{1}{2}\pm I_{2m^{\prime}_{},n^{\prime}_{}}^{}/L_{d}^{}$ are non-negative for every possible value of $m^{\prime}_{}$, $n^{\prime}_{}$, $y_{0}^{}$ and $L_{d}^{}$, implying a positive coupling between electrons and bending vibrons. This can be understood in terms of the deformation potential, inducing a positive coupling to electrons for increasing local areas, which is always the case for bending deformations once compared to the equilibrium straight nanotube configuration. In parallel, we notice that the electron-vibron coupling for bending modes is independent of the graphene mass density but is sensitive to both the length and the circumference of CNT, as well as to the electronic localization.  
In the case of electronic wavefunctions sharply localized near the maximum local strain, the electron-vibron couplings reach the maximum values
\begin{eqnarray}
\label{LambdaMaxBendNum1}
&&\lambda_{m^{\prime}_{},n^{\prime}_{},{\rm max}}^{({\rm Bend}, n=1)}\simeq \frac{40}{m^{\prime 2}_{}}\cdot\frac{L[{\rm \mu m}]}{L_{\perp}^{3}[{\rm nm}]}\nonumber\\
&&\lambda_{m^{\prime}_{},n^{\prime}_{},{\rm max}}^{({\rm Bend}, n\ge 2)}\simeq 10^{-4}_{}\frac{n^{2}_{}+1}{n^{4}_{}}\cdot\frac{L_{\perp}^{}[{\rm nm}]}{L[{\rm \mu m}]}\; .
\end{eqnarray}
In terms of transport measurements on suspended CNT, due to their quadratic coupling bending modes do not produce significant Franck-Condon blockade nor vibrational sidebands at large bias. 
We point out that the apparently large coupling constant of the soft $n=1$ mode is in reality suppressed by the fact that all experimentally realized CNT so far have typical values of $L_{\perp}^{}$ of several nanometers. As an example, the (10,10) armchair CNT is characterized by $L_{\perp}^{}=4.4\, {\rm nm}$, yielding $\lambda_{{\rm max}}^{({\rm Bend}, n=1)}\simeq L[\mu m]/(2m^{\prime 2}_{})$. Only the softest bending vibrons in extremely long and thin freely suspended CNT could ever reach the regime $\lambda_{{\rm max}}^{({\rm Bend}, n=1)}>1$. 

In addition, in experiments the contact-induced tension in the CNT stiffens the low energy dispersion of the $n=1$ mode and reduces its oscillator length and coupling strength.
In particular, the coupling constant becomes 
\begin{equation}
\label{LambdaMaxBendNum2}
\lambda_{\gamma\neq 0}^{({\rm Bend}, n=1)}= \left(\frac{\omega_{n=1, q}^{({\rm Bend})}}{\omega_{n=1, q}^{({\rm Bend,Tens})}}\right)^{2}_{}\cdot \lambda_{\gamma= 0}^{({\rm Bend}, n=1)}
\end{equation}
with the $n=1$ dispersion in Eq.\ (\ref{BendingTension}).

We point out that if the CNT shows buckling in its equilibrium configuration (e.g. via  contact-induced compressions at the two ends), the symmetry argument for bending modes does not hold and a linear electron-vibron coupling appears. In principle, it is straightforward to treat this situation within the approach presented in this paper. However, we refrain from giving explicit results here because of the highly non-universal nature of the effect. 
 
\section{Conclusions}
\label{sec:conclusions}
 
In this paper, we have presented a theory of the intrinsic electron-vibron coupling in carbon-nanotube quantum dots. This intrinsic coupling originates from the electron-phonon interaction of the underlying graphene sheet and is associated with the deformation potential as well as the vibron-induced modulations of the hopping amplitude. We find that for several vibron modes, including those most prominently observed in experiment, this coupling is larger than the previously considered extrinsic coupling based on electrostatic interactions between the nanotube and a nearby gate electrode. 

Our results are consistent with recent experiments in several respects. We find that the most strongly coupled vibron modes are the radial breathing mode and the longitudinal stretching mode. In fact, it is these modes which have been observed in recent transport experiments.\cite{LeRoy,Sapmaz,Huttel1,Huttel2,Leturcq} Moreover, we obtain the largest coupling constants for the longitudinal stretching mode, with the maximal dimensionless coupling exceeding unity, consistent with the recent observation of the Franck-Condon blockade associated with this vibron mode.\cite{Leturcq} 

We also find that quite generally, the electron-vibron coupling is rather sample specific. The coupling depends on the geometric details of the nanotube such as circumference or length and, except for the radial breathing mode, is sensitive to the location of the CNT quantum dot along the vibrating nanotube. 

In very clean CNT, where the quantum dot extends over the entire length of the vibrating nanotube, we find that the electron-vibron coupling is subject to selection rules. 
This significantly suppresses vibronic signatures in the conductance which may be responsible for the absence of vibrational features in some experiments on suspended carbon nanotubes.\cite{Ilani}

In this paper, we have focused on the intrinsic electron-vibron coupling in ideal carbon nanotubes. Realistic samples may also be subject to defects which affect the elastic properties of the carbon nanotubes. Such defects are expected to be particularly significant when they bring the CNT close to an elastic bistability. Another issue which remains an interesting topic for future research concerns consequences of electronic correlations on the electron-vibron coupling. 

Experiments on nanoelectromechanical systems are characterized by an increasing ability to design and tune the sample properties. We expect that the results of the present paper will contribute to extend this ability to include the electron-vibron interaction strengths which may make previously unexplored parameter regimes accessible.

\begin{acknowledgments}
We would like to thank C. Stampfer, R. Leturcq, G. Weick, M. Grifoni, F. Guinea and A. Pikovski for valuable discussions. EM gratefully acknowledges the hospitality of the Aspen Center for Physics. FvO enjoyed the hospitality of the KITP Santa Barbara while this work was completed. This work was supported in part by SPP 1243, DIP and SFB 658. 
\end{acknowledgments}

\section{Appendix}
\label{sec:appendix}

The Euler-Lagrange equations for $u_{x}^{}$,
$u_{y}^{}$ and $h$ out of the Lagrangian density Eq.\ (\ref{elastic}) at the harmonic level are 
\begin{eqnarray}
\label{Euler}
&&\rho_{0}^{}\ddot{u}_{x}^{}=A\partial^{2}_{x}u^{}_{x}+\mu
\partial^{2}_{y}u^{}_{x}+(A-\mu)\partial^{}_{x}\partial^{}_{y}u^{}_{y}
+\frac{A}{r}\partial^{}_{x}h
\nonumber\\
&&\rho_{0}^{}\ddot{u}_{y}^{}=A\partial^{2}_{y}u^{}_{y}+\mu
\partial^{2}_{x}u^{}_{y}+(A-\mu)\partial^{}_{x}\partial^{}_{y}u^{}_{x}
+\frac{\lambda}{r}\partial^{}_{y}h\nonumber\\
&&\rho_{0}^{}\ddot{h}=-A\left(\frac{h}{r^{2}_{}}+\frac{\partial^{}_{x}u_
{x}^{}}{r^{}_{}}\right)-\lambda\frac{\partial^{}_{y}u_{y}^{}}{r^{}_{}}
\\
&&\quad\quad\quad -\kappa\left(\partial^{2}_{x}+\partial^{2}_{y}+\frac{1}{r^{2}_{}}\right)^{2}_{}h\, ,
\nonumber
\end{eqnarray}
with $A=2\mu +\lambda$. 
Imposing periodic boundary conditions along the $x$-direction, we look
for solutions of Eq.\ (\ref{Euler}) of the form $u_{x}^{},\, u_{y}^{},\, h\propto \exp [i nx/r+iqy-i \omega t]$ with $q$ the one-dimensional wavenumber along the nanotube axis.
For $n=0$ we get three equations
yielding the twisting, breathing and stretching eigenmodes
\begin{eqnarray}
\label{Eigenmodes}
&&\rho_{0}^{}\omega^{2}_{}u_{x}^{}=\mu q^{2}_{}u^{}_{x} \nonumber\\
&&\rho_{0}^{}\omega^{2}_{}u_{y}^{}=A\, q^{2}_{}u^{}_{y}-i\lambda\frac
{q}{r}h \\
&&\rho_{0}^{}\omega^{2}_{}h=\frac{A}{r^{2}_{}}h+i\lambda\frac{q}{r}u_{y}^{}+\kappa\left(\frac{1}{r^{2}_{}}-q^{2}_{}\right)^
{2}_{}h\nonumber\; .
\end{eqnarray}

{\it Twist mode}: The first equation in Eq.\ (\ref{Eigenmodes}) yields the twist mode, with   $u_{x}^{}=u\, \sin (qy)$ and $u_{y}^{}=h=0$. Its dispersion
\begin{equation}
\omega_{n=0,q}^{({\rm Twist})}=\sqrt{\frac{\mu}{\rho_{0}^{}}}\, q
\end{equation}
coincides with that of the transverse in-plane phonon branch of graphene.\cite{Mariani}

{\it Breathing mode}: The third equation in Eq.\ (\ref{Eigenmodes}) yields the breathing mode, with finite $h$ and $u_{x}^{}=u_{y}^{}=0$. Due to the CNT curvature, at $q=0$ this distortion yields a finite gap $\omega_{{\rm B}}^{}=\sqrt{(Ar^{2}_{}+\kappa)/(\rho_{0}^{}r^{4}_{})}$. In 
the $qr\ll 1$ regime, we extract $u_{y}^{}$ from the second
equation with $\omega\simeq\omega_{{\rm B}}^{}$
and insert it into the third one, to get the correction
\begin{equation}
\omega_{n=0,q}^{({\rm Breath})}=\omega_{{\rm B}}^{}\left( 1+\beta (qr)^
{2}_{}\right)
\end{equation}
with $\beta =\frac{1}{2(1+\eta )}\left(\frac{\lambda^{2}_{}}{A^{2}_{}(1+\eta)}-2\eta\right)$
and $\eta =\kappa/Ar^{2}_{}\ll 1$. Notice that the group velocity of the breathing mode never becomes negative in realistic CNT.

{\it Stretching mode}: The second equation in Eq.\ (\ref{Eigenmodes}) yields the stretching mode with finite $u_{y}^{}$ and $u_{x}^{}=0$. Obtaining $h$ from the third equation in the regime $\omega
\ll\omega_{{\rm B}}^{}$ and inserting it into the second we obtain
\begin{equation}
\omega_{n=0,q}^{({\rm Stretch})}=\sqrt{\frac{A^{2}_{}-\lambda^{2}_{}}{A\rho_{0}^
{}}}\, q
\end{equation}
in lowest order in $\kappa/ Ar^{2}_{}\ll 1$.

{\it Bending modes}: Further phononic branches at low energy originate from $n\ge 1$. Generalizing the deformations Eq.\ (\ref{Shift})
at finite $q$, from the equations 
(\ref{Euler}) we get the long-wavelength "bending" modes for $qr\ll 1$\\  
\begin{eqnarray}
&& u_{x}^{}=-\frac{u}{n} \sin(nx/r)\sin(qy)\exp [-i\omega
t] \nonumber\\
&& u_{y}^{}=-u \, F_{n}^{}(qr) \cos(nx/r)\cos(qy)\exp [-i\omega
t] \\
&&h=u \, G_{n}^{}(qr) \cos(nx/r)\sin(qy)\exp [-i\omega t]\quad .\nonumber
\end{eqnarray}
In the $\kappa\ll Ar^{2}_{}$ regime the eigenmodes are described by 
\begin{eqnarray}
&&F_{1}^{}(x\ll 1)\simeq x\left(1-\frac{4(\mu +\lambda )}{A}x^{2}_{}\right)\\ 
&&G_{1}^{}(x\ll 1)\simeq 1-\frac{\lambda}{A} x^{2}_{}+
\frac{2(\mu+\lambda)(\mu+2\lambda)}{A^{2}_{}} x^{4}_{}\nonumber
\end{eqnarray}
and by
\begin{eqnarray}
&&F_{n\geq 2}^{}(x\ll 1)\simeq \frac{1}{n^{2}_{}}\, x \\ 
&&G_{n\geq 2}^{}(x\ll 1)\simeq 1-\frac{\kappa}{Ar^{2}_{}}\,\frac{\left( n^{2}_{}-1\right)^{2}_{}}{n^{2}_{}+1}-\frac{1}{n^{2}_{}}\,\frac{\lambda}{A}\, x^{2}_{}\; .\nonumber
\end{eqnarray}
The resulting long wavelength dispersions are
\begin{eqnarray}
&&\omega_{n= 1,qr\ll 1}^{({\rm Bend})}\simeq \left(qr\right)^{2}_{}\sqrt{\frac{2\mu(\mu+\lambda)}{A\rho_{0}^{}r^{2}_{}}}\nonumber\\
&&\omega_{n\geq 2,qr\ll 1}^{({\rm Bend})}\simeq \left(\frac{n_{}^{2}-1}{r^{2}_{}}\right)\sqrt{\frac{\kappa}{\rho_{0}^{}}\cdot\frac{n_{}^{2}}{n^{2}_{}+1}}\times \\
&&\quad\quad\times\left(
1+\left( qr\right)^{2}_{}\frac{2n^{4}_{}A+n^{2}_{}(A-2\lambda)+(A+2\lambda)}{2n^{2}_{}(n^{4}_{}-1)A}\right)\, .\nonumber\\
\nonumber
\label{Bending}
\end{eqnarray}
While the $n\geq 2$ modes are gapped by the bending energy cost, the $n=1$ mode is gapless, with a soft quadratic dispersion predominantly determined by the stretching energy Lam\'e coefficients $\mu$ and $\lambda$.


\begin{thebibliography}{99}
\bibitem{Parameters} Typical parameters for graphene sheets are: $
\kappa =1\,\mathrm{eV}$,  $\mu \simeq 4\lambda =9\,\mathrm{eV}{\rm \AA}^{-2}
_{}$, $\rho_{0}^{}\simeq 5\cdot 10^{-7}_{}\, {\rm kg}\cdot {\rm m}^{-2}_{}$, $g_{1}^{}\simeq 30\, {\rm eV}$ and $g_{2}^{}\simeq 1.5\, {\rm eV}$.
\bibitem{footnote} We note that this assumption may not be valid when considering the radial breathing mode.
\end{thebibliography}
\end{document}